\documentclass[10pt,amsmath,amssymb]{iopart}
\usepackage{graphicx}
\usepackage{iopams}
\begin{document}

\title{Crystallographic disorder and electron scattering on
structural two-level systems in ZrAs$_{1.4}$Se$_{0.5}$}

\author{M Schmidt, T Cichorek\footnote{Permanent address:
Institute of Low Temperature and Structure Research, Polish
Academy of Sciences, 50-950 Wroclaw, Poland.}, R Niewa, A
Schlechte, Y Prots, F~Steglich and R Kniep}

\address{Max Planck Institute for Chemical Physics of Solids, Noethnitzer Str. 40, 01187~Dresden, Germany}

\date{\today}

\begin{abstract}

Single crystals of ZrAs$_{1.4}$Se$_{0.5}$ (PbFCl type structure)
were grown by chemical vapour transport. While their thermodynamic
and transport properties are typical for ordinary metals, the
electrical resistivity exhibits a shallow minimum at low
temperatures. Application of strong magnetic fields does not
influence this anomaly. The minimum of the resistivity in
ZrAs$_{1.4}$Se$_{0.5}$ apparently originates from interaction
between the conduction electrons and  structural two-level
systems. Significant disorder in the As-Se substructure is
inferred from X-ray diffraction and electron microprobe studies.

\end{abstract}


\section{Introduction}
\label{sec:intro}

Several thermal, electrical and magnetic properties of ordinary
metals in the presence of dilute magnetic impurities appear to be
anomalous with respect to the properties observed in the presence
of nonmagnetic impurities. For instance, the electrical
resistivity $\rho$(\textit{T}) of an ordinary metal containing
nonmagnetic impurities decreases monotonically with decreasing
temperature and becomes constant at low temperatures. On the
contrary, the resistivity of a normal metal in the presence of
dilute magnetic impurities exhibits a rather shallow minimum at
low temperatures \cite{meissner, kondo1}. This resistivity anomaly
reflects an unusual nature of the scattering of conduction
electrons from an isolated impurity spin. As originally suggested
by Cochrane \textit{et al}. \cite{coch}, qualitatively different
scattering of conduction electrons from nonmagnetic and magnetic
impurities can be lifted in metals with structural two-level
systems (TLS), i.e., tunneling defects with two energy levels. The
simplest realization of a tunneling center is believed to be an
atom that quantum-mechanically tunnels between two metastable
states of the double-well potential. Indeed, under some
circumstances, internal degrees of freedom of a tunneling center
may be mapped to internal degrees of freedom of the magnetic
impurity embedded in the Fermi sea and hence an additional,
\textit{T}-dependent resistivity, being hardly affected by a
strong magnetic field, may occur at low temperatures \cite{cox}.

Interactions between the conduction electrons and structural TLS
appear to be remarkably strong in certain arsenide selenides. For
example, the resistivity of diamagnetic ThAsSe frequently displays
a logarithmic correction below around 20~K, which is affected by
neither strong magnetic fields nor high hydrostatic pressures
\cite{cich1}. The existence of tunneling centers in single
crystals of ThAsSe is reflected by, e.g., a glassy-type
temperature dependence of both the thermal conductivity and
specific heat at \textit{T}$\lesssim$1 K \cite{cich1, cich3}. For
UAsSe and its derivatives, a strongly sample-dependent upturn in
$\rho$(\textit{T}) is observed deep in the ferromagnetic state
\cite{cich2}. Whereas an influence of structural TLS on the charge
transport in actinide-based arsenide selenides is well
established, very little is known on the nature of tunneling
centers in these materials \cite{henk1, henk2}.

Guided by remarkably similar physical properties of ThAsSe and
UAsSe, we have grown single crystals of an isostructural Zr-based
compound to verify our hypothesis on an interaction between the
conduction electrons and TLS in these materials. In this paper we
report on basic physical properties of ZrAs$_{1.4}$Se$_{0.5}$,
compare them with those of ThAsSe, and discuss some
crystallochemical aspects of the As-Se substructure being
indicative of structural disorder in layered arsenide selenides.

\section{Experimental details}

Single crystals of the ternary phase were grown by chemical vapour
transport from a pre-reacted substrate. The micro-crystalline
powder had been synthesized by reaction of the elements (Zr 99.8\%
Chempur, As 99.9999\% Chempur, Se 99.999\% Chempur) in a molar
ratio of 1:1:1. The elements were placed in an evacuated quartz
ampoule and a glassy carbon crucible was used to prevent a
possible reaction of zirconium with silica. The reaction was
conducted by a gradual increase in temperature from 473 K to 1173
K over 14 days followed by a subsequent treatment at the final
temperature for 14 days more. For the crystal growth experiments,
iodine (Chempur 99.999\%) served as a transport agent (1
mg/cm$^{3}$). The deposition of crystals were realized
exothermically from \textit{T}$_{1}$ = 1123 K to \textit{T}$_{2}$
= 1223 K over a distance of 10 cm (diameter of the ampoule:
20~mm). The as-grown crystals show metallic lustre. All presented
data were obtained on one single crystal or parts of it. The
specimen investigated has a size of 2 mm $\times$ 0.5 mm $\times$
0.3 mm and a mass of 2 mg (figure 1).

The chemical composition of the crystal was initially examined by
scanning electron microscopy with energy dispersive X-ray
microanalyses utilizing a Philips XL30 scanning electron
microscope. The precise composition was determined on polished
surfaces using electron-probe microanalyse. The investigations
were done with a wavelength dispersive system (WDX) Cameca SX100.
Elemental standards for zirconium, arsenic and selenium were used.

The quality of the crystal was checked with Laue diffractograms.
Room-temperature X-ray diffraction intensity data were collected
on a STOE IPDS diffractometer using graphite monochromated Ag\
K$\alpha$ radiation (above the absorption threshold of the Zr\
K-edge). Accurate unit cell parameters were determined from the
ground crystal on a STOE Stadi-mp diffractometer using Cu\
K$\alpha$$_{1}$ radiation after all the measurements had been
performed. The structure was solved and refined with the SHELX
package of programs \cite{sheld1, sheld2}. A numerical absorption
correction of the intensity data was carried out for the optimized
shape of the crystal by using STOE X-RED and X-SHAPE programs.

\begin{figure}\centerline
{\includegraphics[width=7.5cm,keepaspectratio]{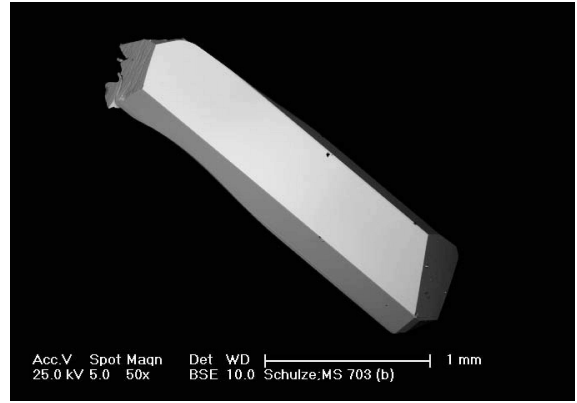}}
\caption{\label{fig:1} Single crystal of ZrAs$_{1.40}$Se$_{0.50}$
investigated in this work.}
\end{figure}

The electrical resistivity was investigated along the \textit{a}
axis by a conventional four-point ac method. Experiments were
performed in zero and applied magnetic field of 9~T down to 2~K.
Electrical contacts were made by spot welding 25 $\mu$m gold wires
to the crystal. The heat capacity was determined with the aid of
the thermal-relaxation technique utilizing a commercial
microcalorimeter (Quantum Design, model PPMS). The dc magnetic
susceptibility was measured using a SQUID magnetometer (Quantum
Design, model MPMS).

\section{Results}

\subsection{Crystal structure and chemical composition}

\begin{figure}\centerline
{\includegraphics[width=10.0cm,keepaspectratio]{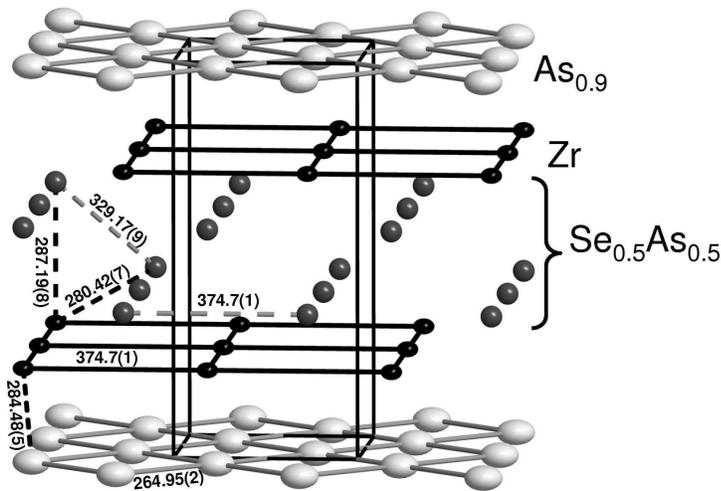}}
\caption{\label{fig:1}~PbFCl type crystal structure of
ZrAs$_{0.90(1)}$(Se$_{0.50(1)}$As$_{0.50}$) =
ZrAs$_{1.40}$Se$_{0.50}$. Note the 90\% occupation of the As site
and the anomalous displacement parameters. Distances are given in
pm.}
\end{figure}

\begin{table}
  \caption{Positional and displacement parameters (10$^{-4}$ pm$^{2}$) for ZrAs$_{1.40}$Se$_{0.50}$.}\label{1}
  \begin{indented}\item[]
  \begin{tabular}{l c c c c c c l}
atom & site & \textit{x} & \textit{y} & \textit{z} &
\textit{U}$_{11}$=\textit{U}$_{22}$ & \textit{U}$_{33}$ & Occ.
\\\hline
  Zr & 2\textit{c} & 1/4 & 1/4 & 0.26521(6) & 0.0050(2) & 0.0031(2) & 1\\
  As & 2\textit{a} & 3/4 & 1/4 & 0 & 0.0111(3) & 0.0034(3) & 0.906(6)\\
  Se/As & 2\textit{c} & 1/4 & 1/4 & 0.62099(7) & 0.0044(2) & 0.0041(3) & 0.501(6)/0.499\\\hline
  \end{tabular}
  \begin{tabular}{l }
    \textit{U}$_{12}$=\textit{U}$_{13}$=\textit{U}$_{23}$=0\\
  \end{tabular}
  \end{indented}
\end{table}

\begin{table}
  \caption{Crystallographic data for ZrAs$_{1.40}$Se$_{0.50}$.}\label{2}
  \begin{indented}\item[]
  \begin{tabular}{l l}
  \hline
     Crystal size (mm$^{3}$) & 0.035$\times$0.075$\times$0.090\\
     Space group & \textit{P}4/\textit{nmm} (No. 129)\\
     Lattice parameters (pm) & \textit{a} = 374.69(1)\\
      & \textit{c} = 807.16(2)\\
      Cell volume (10$^{6}$ pm$^{3}$), \textit{Z} & 113.316(4), 2\\
     Image plate distance (mm) & 60\\
     $\rho$ range (deg), $\Delta\rho$ (deg) & 200, 1\\
    2$\theta$ range (deg) & 2$\theta<$ 55.8\\
   \textit{hkl} range & -6$\leq$\textit{h}$\leq$6\\
    & -6$\leq$\textit{k}$\leq$6\\
    & -13$\leq$\textit{l}$\leq$13\\
    No.\ of measured reflections, \textit{R}$_{int}$ & 2183, 0.038\\
    No.\ of independent reflections & 200\\
    No.\ of refined parameters & 12\\
    Program & SHELX97 [11]\\
   \textit{R}$_{\rm gt}$(\textit{F}), \textit{R}$_{\rm all}$(\textit{F}) & 0.019, 0.042\\
    GooF & 0.882\\
    Extinction coefficient & 0.28(1)\\
    Largest peaks in difference electron density & 0.95, -2.49\\
   $\Delta\rho_{\rm max}$, $\Delta\rho_{\rm min}$ (10$^{-6}$ pm$^{-3}$) & \\\hline
  \end{tabular}
  \end{indented}
\end{table}

The obtained single crystal diffraction data set could be indexed
on the basis of a tetragonal unit cell with \textit{a} = 374.17(7)
pm, \textit{c} = 811.9(2) pm. A precise unit cell refinement with
Guinier technique resulted in \textit{a} = 374.69(1) pm,
\textit{c} = 807.16(2) pm. The extinction conditions led to the
space groups \textit{P}4/\textit{n} and \textit{P}4/\textit{nmm}.
Refinements in both space groups resulted in qualitatively
identical structure models, therefore the higher symmetry space
group \textit{P}4/\textit{nmm} was chosen for the detailed
structure analysis.

After an initial refinement of an ordered ZrSiS type structure As
was successively introduced on the Se site leading to a structure
model with the composition ZrAs(Se$_{0.60(1)}$As$_{0.40}$) and
reliability factors of \textit{R}1/\textit{wR}2 = 0.030/0.074. On
refinement the As site occupancy dropped to 0.906(6), while the
Se/As ratio on the Se site adjusted to 0.501(6)/0.499 and the
reliability factors dropped to \textit{R}1/\textit{wR}2 =
0.019/0.042 in a stable refinement. The resulting composition from
X-ray diffraction analysis is
ZrAs$_{0.90(1)}$(Se$_{0.50(1)}$As$_{0.50}$) =
ZrAs$_{1.40}$Se$_{0.50}$  (figure 2). The anisotropic displacement
parameters are nearly not affected by these operations. This
indicates an excellent structural model intensity description,
which is additionally in excellent agreement with WDX analyses.
Tables 1 and 2 gather structure determination and crystallographic
data. A refinement of intensity data obtained with Mo\ K$\alpha$
radiation (below the absorption threshold of the Zr\ K-edge) did
not allow any refinement of the ratio As/Se.

ZrAs$_{0.90(1)}$(Se$_{0.50(1)}$As$_{0.50}$) =
ZrAs$_{1.40}$Se$_{0.50}$ crystallizes in the UPS branch
(substitution variant of the Fe$_{2}$As type, often referred to as
ZrSiS type) of the PbFCl type structure characterized by a larger
ratio \textit{c}/\textit{a} $\geq$ 2.0 together with a large
fractional parameter of the metal atom (Zr) \textit{z} $\geq$ 2.2
as compared to the BiOI type \cite{fla}. In this structure type
typically the larger anions occupy the 2\textit{c} site and the
smaller anions the 2\textit{a} site. Both anion sites might be
susceptible to substitution with the respective other species or a
third element with the restriction of atomic/ionic radii fit
\cite{wang}. For the ZrSiS branch of MXY compounds appreciable
X--X bonding in the densely packed quadratic layer of X is
commonly discussed, while Y--Y bonding is typically moderate.

Refinements of As/Se ratios based on X-ray diffraction data
presents a problem due to the difference of only one electron for
those elements. Additionally, a scaling problem for refinement of
the As-site occupation arises. Still, the obtained data are in
excellent agreement with the WDX analysis results
(ZrAs$_{1.40(1)}$Se$_{0.50(1)}$ vs 34.69(4) At \% Zr, 48.51(5) At
\% As, and 16.80(3) At \% Se, i.e., ZrAs$_{1.398}$Se$_{0.484}$)
for the identical crystal supporting the composition and
crystallographic disorder model. For the As-rich phase in the
system Zr-As-Se disorder with mixed occupation of the 2\textit{c}
site by Se and As was earlier discussed by Barthelat \textit{et
al} \cite{bar1, bar2} leading to a composition of
ZrAs$_{0.78}$(As$_{0.37}$Se$_{0.58}$) = ZrAs$_{1.15}$Se$_{0.58}$.

\subsection{Physical properties}

The temperature dependence of the specific heat
\textit{c$_{p}$}(\textit{T}) of the ZrAs$_{1.4}$Se$_{0.5}$ single
crystal is depicted in figure~3. The heat-capacity examination has
not shown any phase transition. The low-temperature
\textit{c$_{p}$}(\textit{T}) data are presented as
\textit{c$_{p}$}/\textit{T} vs \textit{T}$^{2}$ in the inset of
figure~3. From the straight line found for \textit{T} $\leqslant$
9\ K one obtains the Sommerfeld coeffcient $\gamma$ =
1.7($\pm$0.2) mJ\ K$^{-2}$\ mol$^{-1}$ and the slope $\beta$ =
1.42($\pm$0.05)$\times$10$^{-4}$ J\ K$^{-4}$\ mol$^{-1}$, which
yields a Debye temperature $\Theta_{\rm D}$ = 345\ K.

The absence of any phase transition is also inferred from the
magnetic susceptibility $\chi$ of ZrAs$_{1.4}$Se$_{0.5}$ which is
hardly temperature-dependent and negative in the temperature range
2--300~K, i.e., typical for a diamagnetic material. At 300~K and
in field of 4~T applied along the \textit{a} axis, a magnetic
susceptibility of $-$5.5$\times$10$^{-4}$\ emu/g was found.
Unfortunately, the small mass of the crystal prevented us from a
more detailed study of $\chi$(\textit{T}) in lower fields.

\begin{figure}\centerline
{\includegraphics[width=12.0cm,keepaspectratio]{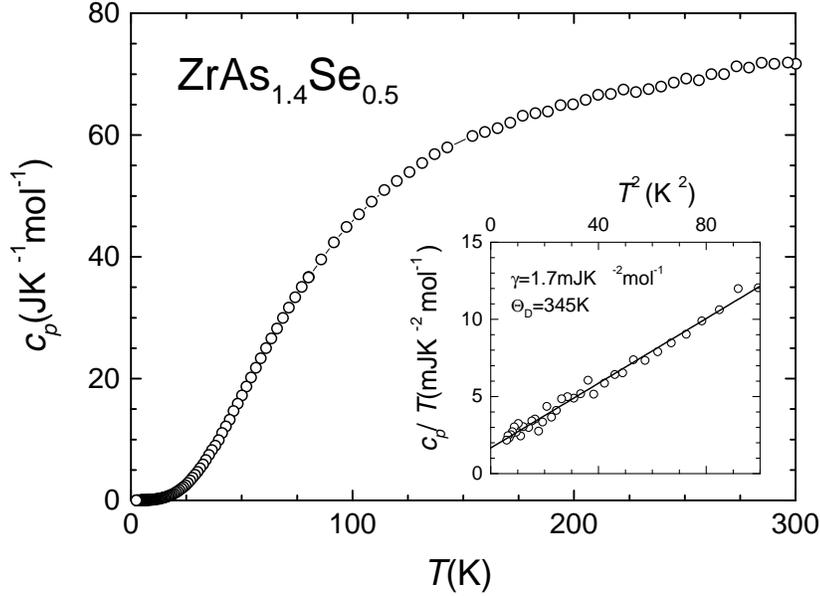}}
\caption{\label{fig:2} Temperature dependence of the heat capacity
for a single crystal of ZrAs$_{1.4}$Se$_{0.5}$. Inset: The
low-temperature specific heat, as \textit{c$\rm_{p}$}/\textit{T}
vs \textit{T}$^{2}$, between 1.8 and 10~K. The solid curve
represents a $\gamma$\textit{T} + $\beta$\textit{T}$^{3}$
dependence with $\beta$=3$\times$1944/$\Theta_{\rm D}^{3}$ in
units of J\ K$^{-4}$\ mol$^{-1}$ and we have neglected defects in
the 2\textit{a} anionic site.}
\end{figure}

The temperature dependence of the electrical resistivity of
ZrAs$_{1.4}$Se$_{0.5}$ shows a well-defined metallic behaviour, as
presented in figure~4. Indeed, the $\rho$(\textit{T}) data along
the \textit{a} axis are accounted for in terms of the
Bloch-Gr\"{u}neisen model with  a slight correction due to
\textit{s}--\textit{d} interband electron scattering. This is
especially true at higher temperatures where a nonlinear
\textit{T} dependence is observed. A least squares fit according
to a generalized Bloch-Gr\"{u}neisen-Mott relation with power
\textit{n} = 3 \cite{grimvall}:

\begin{equation}\label{1}
  \rho(T)=\rho_{0}+\textit{C}\left(\frac{T}{\Theta_{\rm D}^{\rm R}}\right)^{n}\int_{0}^{\Theta\rm_{D}^{R}/T}\frac{x^{n}dx}{(e^{x}-1)(1-e^{-x})}+KT^{3}
\end{equation}
yields the residual resistivity $\rho_{0}$ = 140 $\mu\Omega$cm and
material dependent constants $\Theta_{\rm D}^{\rm R}$ = 303 K,
\textit{C} = 54.5 $\mu\Omega$cm, and \textit{K} =
$-$1.77$\times$10$^{-7}$ $\mu\Omega$cm\ K$^{-3}$ (solid line in
figure 4).

However, a closer inspection of our low-temperature
$\rho$(\textit{T}) results for ZrAs$_{1.4}$Se$_{0.5}$ reveals an
additional resistivity. Remarkably, this extra
\textit{T}-dependent term is not influenced by strong magnetic
fields. Details are depicted in the inset of figure\ 4. Here we
have plotted the relative change of the resistivity normalized to
the corresponding value at 2~K obtained in \textit{B} = 0 and 9~T.
It is noticeable that the very similar correction to the
low-\textit{T} resistivity has been detected in an overwhelming
majority of the Zr/As/Se single crystals.

\begin{figure}\centerline
{\includegraphics[width=12.0cm,keepaspectratio]{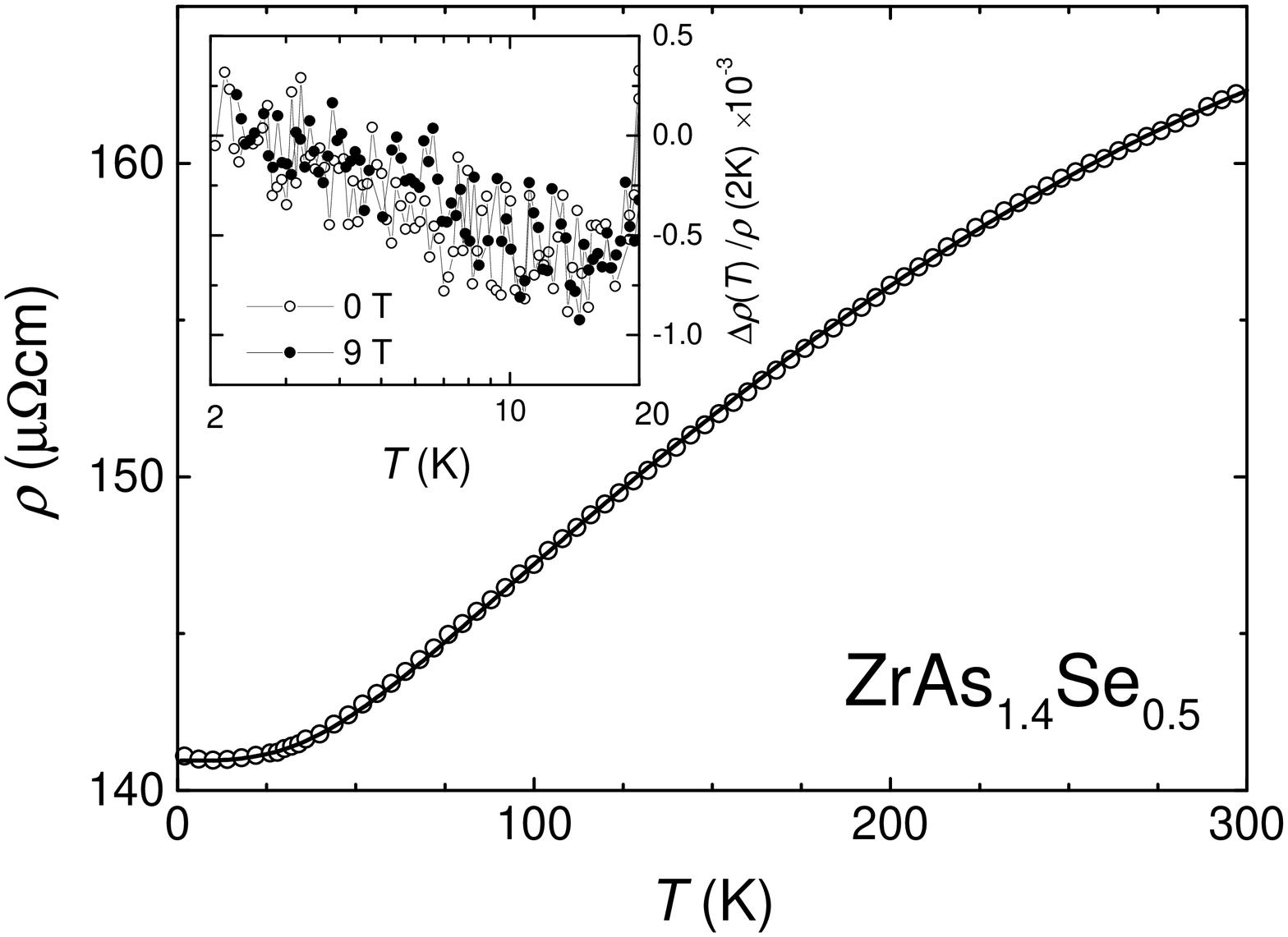}}
\caption{\label{fig:3}The electrical resistivity for a single
crystal of ZrAs$_{1.4}$Se$_{0.5}$ as a function of temperature.
The solid line is a fit as described in the text. Inset: The
relative change of the low-temperature resistivity normalized to
the corresponding value at 2\ K obtained in zero field and 9\ T.}
\end{figure}

\section{Discussion and conclusions}

Diamagnetic ZrAs$_{1.4}$Se$_{0.5}$ exhibits features typical of an
ordinary metal. The observed value of the Sommerfeld coefficient
$\gamma$ = 1.7($\pm$0.2) mJK$^{-2}$mol$^{-1}$ is comparable to the
electronic term in the specific heat of such metallic elements as,
e.g., In, Sn or Hg. On the other hand, the residual resistivity
ratio as small as 1.15 resembles rather some alloys (e.g.,
manganin) than good metals and may be treated as an approximate
indicator of significant disorder in the ZrAs$_{1.4}$Se$_{0.5}$
crystals. This is further suggested by a large value of $\rho_{0}$
= 140 $\mu\Omega$cm. Most importantly, however, a significant
disorder in our samples is directly inferred from the results of
both the electron-probe microanalysis and the x-ray diffraction
studies. In fact, since any deviation from the 1:1:1 stoichiometry
of the system crystallizing in the PbFCl type structure always
results in structural disorder, the number of imperfections has to
be large in specimens with the chemical composition of 1:1.4:0.5.
This general information is resolved by the results of the
structure refinement indicative of one anionic site mixed occupied
by 50\% As and 50\% Se and the 10\% defects on the second anionic
site.

A minimum resistivity at around 12\ K is the most intriguing
physical property of ZrAs$_{1.4}$Se$_{0.5}$. In fact, since
virtually the same $\Delta\rho(\textit{T})/\rho$(2\ K) data were
obtained at both \textit{B} = 0 and 9\ T, a magnetic Kondo effect
is very unlikely in this material \cite{magnetic}. Especially that
a nonmagnetic anomaly of similar amplitude has been observed in
all the single crystals of the related system ThAsSe. In this
material electron scattering on magnetic impurities had been
unambiguously ruled out \cite{cich1}. Therefore, striking
similarity in the low-\textit{T} behaviour of $\rho(\textit{T})$
for ThAsSe and ZrAs$_{1.4}$Se$_{0.5}$ allows us to consider the
same origin of an electron scattering in both arsenide selenides
that we relate to a presence of tunneling centers in the As--Se
substructure. Indeed, although the interatomic distances between
the 2\textit{c} sites are comparably large, arsenic and selenium
atoms may be involved in a homopolar-to-heteropolar bond
transformation, as discussed for As$_{2}$Se$_{3}$ \cite{Li} and
As$_{2}$S$_{3}$ \cite{Uchino}. Such a possibility is also
plausible for ThAsSe, since in some specimens of this system a
large As/Se $\simeq$ 1.5 content ratio was recently observed
\cite{burkhardt}.

By contrast to ZrAs$_{1.4}$Se$_{0.5}$, the Th-based compound
displays a negative temperature coefficient of the
\textit{ab}-plane resistivity at temperatures higher than 65~K.
(The same holds true for UAsSe in the paramagnetic state). As
speculated in Ref.\ \cite{schoe}, this high-temperature increase
of $\rho$(\textit{T}) may be caused by a gradual formation of
covalently bonded dimers (As--As)$^{4-}$ that influences the
electronic structure. This scenario, being based on quite a
general tendency of As ions towards homoatomic bonding in the
quadratic net, was recently supported by the results of electron
diffraction in ThAsSe \cite{withers}. An As--As dimerization is,
to a certain extent, also possible in ZrAs$_{1.4}$Se$_{0.5}$,
although its influence on the charge transport is not observed (cf
figure 3). In fact, enhanced displacement parameters
\textit{U}$_{11}$ = \textit{U}$_{22}$ $\gg$ \textit{U}$_{33}$ for
As, listed in Table 1, indicate the formation of As--As covalent
bonds leading to diverse possible As$_{n}$ anionic species and
thus to static displacement from the ideal As position in favour
of dynamic vibration. Average distances As--As for the ideal
crystallographic position in the quadratic net from the structure
refinement are \textit{d}(As--As) = 264.95(1) pm. For comparison
in the grey modification of arsenic As has three nearest
neighbours at 252 pm and three further neighbours at 312 pm
\cite{schiferl}. As$_{4}$ molecules in the gas phase exhibit a
distance of 243.5\ pm \cite{morino}. For the As species within the
2\textit{c} site disordered with Se species such formation of
dimers or larger units As$_{n}$ is unlikely, due to the
considerably longer distance of \textit{d}(As--As) $\geq$ 329 pm
in ZrAs$_{1.40(1)}$Se$_{0.50(1)}$.

Though the formation of covalently bonded units frequently occurs
in pnictide chalcogenides crystallizing in the layered PbFCl type
structure, its direct impact on excitations of the electron gas in
metallic arsenide selenides at \textit{T} $\lesssim$ 20\ K is
dubious because: first, an (As--As)$^{4-}$ dimerization is already
completed at much higher temperatures, as inferred from virtually
identical electron diffraction patterns at 30 and 100~K in ThAsSe
\cite{withers}. Second, the dimerization shows a marked
sensitivity to the interatomic distances being easy influenced by
an external pressure. However, the application of high pressure
alters the resistivity of ThAsSe only at high temperatures: At
1.88~GPa the maximum of the resistivity at 65~K is suppressed,
whereas the low-\textit{T} term is completely unchanged
\cite{cich1}. Finally, ZrAs$_{1.4}$Se$_{0.5}$ and ThAsSe show
\textit{qualitatively} different $\rho$(\textit{T}) dependencies
at higher temperatures, although in both diamagnets remarkably
similar low-\textit{T} anomalies in the electrical resistivity are
observed. Nevertheless, it is conceivable that some low-energy
excitations of singular (As--As)$^{4-}$ dimers (but not
dimerization itself) create tunneling centers, being responsible
for the shallow minimum in $\rho$(\textit{T}) of both arsenide
selenides at low temperatures.

As far as the high temperature $\rho$(\textit{T}) data for
ZrAs$_{1.4}$Se$_{0.5}$ are concerned, a well-defined metallic
character of the resistivity for a ThPS system should be quoted
\cite{wawryk}. The ferromagnetic UPS counterpart shows all the
features characteristic of UAsSe, i.e., a strongly
sample-dependent upturn in $\rho$(\textit{T}) deep in the
ferromagnetic state and the negative temperature coeffcient of the
resistivity in the paramagnetic state  \cite{cich2}. However, a
low-\textit{T} increase of $\rho$(\textit{T}) was detected in none
of various ThPS samples so far \cite{wawryk}.

To summarize, we have investigated one single crystal of metallic
ZrAs$_{1.40}$Se$_{0.50}$ which shows a significant atomic disorder
in the As-Se substructure. At temperatures below 12\ K,
ZrAs$_{1.40}$Se$_{0.50}$ displays an unusual, magnetic field
independent increase of the electrical resistivity upon cooling.
Since remarkably similar correction to low-temperature
$\rho$(\textit{T}) had been observed in the related diamagnet
ThAsSe, an anomalous scattering mechanism in the isostructural
Zr-based system appears to be also caused by nonmagnetic
interactions between the conduction electrons and structural
two-level systems. Results of X-ray diffraction and electron
microprobe investigations point out that a formation of tunneling
centers might be triggered off by empty places in the As
(2\textit{a}) layers or/and the mixed occupation of the
2\textit{c} sites by arsenic and selenium.

\section{Acknowledgments}

We would like to thank U. Burkhardt and K. Schulze for their help
with WDX, and R. Cardoso-Gil for assistance with powder X-ray
diffraction. We also want to thank W. Schnelle for a critical
reading of the manuscript.

\end{document}